\documentclass[fleqn,usenatbib]{mnras}

\usepackage{newtxtext,newtxmath}

\usepackage[T1]{fontenc}

\DeclareRobustCommand{\VAN}[3]{#2}
\let\VANthebibliography\thebibliography
\def\thebibliography{\DeclareRobustCommand{\VAN}[3]{##3}\VANthebibliography}

\let\vec\mathbf
\newcommand{\kms}{\,km\,s$^{-1}$}

\usepackage{graphicx}	
\usepackage{amsmath}	

\title[A VLBI SETI Observation of Kepler-111 b]{An Interferometric SETI Observation of Kepler-111~b}

\author[Kelvin Wandia et al.]{
Kelvin Wandia$^{1}$,\thanks{E-mail: kelvin.wandia@manchester.ac.uk}
Michael A. Garrett$^{1,2}$,
Jack F. Radcliffe$^{1,3}$,
Simon T. Garrington$^{1}$,
James Fawcett$^{1}$,
\newauthor
Vishal Gajjar$^{4}$,
David H. E. MacMahon$^4$,
Eskil Varenius$^5$,
Robert M. Campbell$^6$,
Zsolt Paragi$^6$,
\newauthor
Andrew P. V. Siemion$^{1,4,7,8}$
\\
$^{1}$ Jodrell Bank Centre for Astrophysics (JBCA), Department of Physics \& Astronomy, Alan Turing Building, The University of Manchester, M13 9PL, UK\\
$^{2}$ Leiden Observatory, Leiden University, PO Box 9513, 2300 RA Leiden, The Netherlands\\
$^{3}$ Department of Physics, University of Pretoria, Lynnwood Road, Hatfield, Pretoria, 0083, South Africa \\
$^{4}$ Breakthrough Listen, University of California, Berkeley, CA 94720, USA\\
$^{5}$ Department of Space, Earth and Environment, Chalmers University of Technology, Onsala Space Observatory, 439 92 Onsala, Sweden \\
$^{6}$ Joint Institute for VLBI in Europe, Oude Hoogeveensedijk 4, 7991 PD, Dwingeloo, The Netherlands \\
$^{7}$ SETI Institute, Mountain View, CA 94043, USA\\
$^{8}$ University of Malta, Institute of Space Sciences and Astronomy, Malta
}

\date{13 April 2023}

\pubyear{2015}

\begin{document}
\label{firstpage}
\pagerange{\pageref{firstpage}--\pageref{lastpage}}
\maketitle

\begin{abstract}
The application of Very Long Baseline Interferometry (VLBI) to the Search for Extraterrestrial Intelligence (SETI) has been limited to date,  despite the technique offering many advantages over traditional single-dish SETI observations. In order to further develop interferometry for SETI, we used the European VLBI Network (EVN) at $21$~cm to observe potential secondary phase calibrators in the Kepler field. Unfortunately, no secondary calibrators were detected. However, a VLBA primary calibrator in the field, J1926+4441, offset only  $\sim1.88\arcmin$ from a nearby exoplanet Kepler-111~b, was correlated with high temporal $\left(0.25 \ \rm{s}\right)$ and spectral $\left(16384 \times 488\ \rm{Hz \ channels}\right)$ resolution.  During the analysis of the high-resolution data, we identified a spectral feature that was present in both the auto and cross-correlation data with a central frequency of $1420.424\pm0.0002$ MHz and a width of 0.25 MHz. We demonstrate  that the feature in the cross-correlations is an artefact in the data, associated with a significant increase in each telescope's noise figure due to the presence of \ion{H}{i} in the beam. This would typically go unnoticed in data correlated with standard spectral resolution. We flag (excluded from the subsequent analysis) these channels and phase rotate the data to the location of Kepler-111~b  aided by the GAIA catalogue and search for signals with $\rm{SNR}>7$. At the time of our observations, we detect no transmitters with an Equivalent Isotropically Radiated Power (EIRP) > $\sim4\times10^{15}$ W.

\end{abstract}




\section{Introduction}

The origins of the search for extraterrestrial intelligence (SETI) in the radio band can be traced back to Frank Drake \citep{Drake1961}
who made the first targeted SETI searches using the 85-foot Tatel telescope at Greenbank, West Virginia. This groundbreaking work followed a seminal paper by \citet{COCCONI1959}; they argued that the frequency range centred on the neutral hydrogen (\ion{H}{i}) spectral line was an optimal region of the electromagnetic spectrum to search for signals intended for interstellar communications. Even though the original observational setup by \citet{Drake1961} is rudimentary compared to modern standards, the basic tenets of SETI have remained largely unchanged, with large single dishes searching for narrow-band (coherent) radio emission from nearby stars \citep[e.g.][]{Price2020}.
Interferometric SETI using relatively short baselines ($< 11\ \rm{km}$) has been conducted before, especially using the Very Large Array (VLA) and the Murchison Widefield Array (MWA). For example, \cite{Gray2017} searched for signals from M31 and M33 with the VLA and \citep{Chenoa2020} did a survey of the Vela region using the MWA. Until recently, the application of long baseline interferometry to SETI has not been well explored, partly because of the limited capabilities of correlators to generate high spectral resolution data over a wide frequency range. The data rates generated are also large -- these scale as $N(N-1)$ for an array of $N$ antennas compared to just a single dish. As a result, beam-forming techniques have been widely used for SETI surveys using telescope arrays, in particular at the Allen Telescope Array \citep[ATA; ][]{ATA} and more recently using MeerKAT in South Africa \citep{MeerKAT2016} by Breakthrough Listen \citep{BLMeerKAT}. The advent of software correlators which are capable of correlating data at very high spectral and temporal resolution such as the EVN software correlator at JIVE, the Super FX Correlator \citep[SFXC; ][]{Keimpema2015} and Distributed FX \citep[DiFX; ][]{Deller2007} at the VLBA, now make long baseline (non-beam formed) SETI observations an interesting proposition.

The first Very Long Baseline Interferometry (VLBI) SETI experiment was made by \citet{Rampadarath2012} using the Australian Long Baseline Array (LBA) observing Gliese 581 in the frequency range 1230–1544 MHz. They constrained the power of an isotropic emitter to an upper limit of $7\ \rm{MW}\ {Hz^{-1}}$ and also highlighted the advantages of VLBI SETI. The key advantage they identified was that a VLBI array is less susceptible to the effects of Radio Frequency Interference (RFI), a major source of false positives that can sometimes mimic narrow-band, Doppler drifting SETI signals e.g. Breakthrough Listen candidate 1 \citep[BLC1;][]{Smith2021,Sheikh2021}. The advantages of an interferometer are that local sources of RFI do not correlate on long baselines and other mutual sources of RFI (e.g. satellites) de-correlate rapidly if they are located outside the interferometer’s typically narrow field of view. In this respect, longer baselines suppress common RFI sources more than short baselines. 

\citet{Garrett2018} presented other advantages of an interferometric approach to SETI. They noted that the independent detection of candidate signals on multiple telescope baselines adds a level of robustness to detecting a signal that is simply not possible using a single dish. This redundancy could be extremely important in terms of the verification of detections associated with faint, transitory signals.  In addition, since interferometer data can be transformed to the image domain, the 2-D location of a bonafide SETI signal should be invariant with time (unless the signal is located nearby) – this is a fundamental constraint when potentially almost everything else associated with a SETI signal could be changing (e.g. frequency drift due to Doppler accelerations, temporal variability). Related to this is the ability of VLBI to pinpoint the location of an extra-terrestrial signal with (sub-)milliarcsecond precision, and to monitor any motion of the transmitter via multi-epoch observations. This may be crucial in understanding the characteristics of the platform on which the transmitter is fixed (e.g. the orbit determination of any host planet, proper motion of a free-flying platform). We also note that in addition to narrowband signals, a new category of broadband millisecond width transient signals also exists \citep[e.g.][]{Gajjar2022AI}. 
The ability to localize these so-called Fast Radio Bursts (FRBs) has recently been demonstrated by VLBI  \citep[see][]{Nimo2022}, and these techniques can also be applied to SETI candidate signals that are also transient in nature. Finally, the requirement from SETI for high time and frequency resolution is also well matched to wide-field VLBI observations that preserve sensitivity to SETI signals across the full telescope primary beam. This is important, as it maximises serendipity in making a possible SETI detection while beam formed arrays use narrow pencil beams that only cover a fraction of the primary beam of the individual antennas \citep[e.g.][]{GajjarWhitePaper2019,Czech2021}.

To further advance the use of VLBI for SETI, we were successful in obtaining observations of the Kepler field \citep[][]{2006AAS...20921008B,2010ApJ...713L.109B} using the European VLBI Network (EVN). The main goals were to identify secondary phase calibrators in the field and also analyse high spectral and time resolution data of the VLBA calibrator J1926+4441. J1926+4441 is a flat-spectrum radio source ($\alpha>-0.5$) \citep{2007ApJS..171...61H}. This source lies only $1.88\arcmin$ from Kepler-111, a star which is orbited by a (possibly rocky) exoplanet, Kepler-111~b \citep{Rowe}. $e$-MERLIN observations were also made in order to further understand a narrow-band feature in the EVN auto and cross-correlation data. In Section 2 of this paper, we describe the EVN and $e$-MERLIN observations, and the data analysis. In Section 3 we present our results, including the analysis of the unexpected spectral feature in the auto and cross-correlation data. In Section 4 we draw our main conclusions.

\section{Observations and Data Analysis}

The original primary goal of the EVN short observations was to detect secondary VLBI phase calibrators in the Kepler field, in order to support follow-up VLBI SETI surveys of this area of sky. The motivation was based on the fact that this field contains many exoplanets of particular interest since they are well characterised in terms of their mass and radius \citep{https://doi.org/10.26133/nea12}. These and other exoplanet candidates in the field could be interesting targets for VLBI SETI in the future. Secondary calibrators that lie closer to the exoplanet permit better phase corrections to be applied to the VLBI data, substantially improving the coherence of the data compared to that referenced via a more distant (primary) calibrator. Another goal of our observations was to develop and extend the VLBI SETI technique \citep[see][]{Garrett2018} by correlating a single scan on one of the primary calibrators (J1926+4441) with high spectral and temporal resolution. A known exoplanet, Kepler-111~b lies within the field of view centred on J1926+4441 and this permitted a search for narrow-band signals from this system.

\subsection{Phase calibrator target selection}

\begin{figure}
    \includegraphics[width=\columnwidth]{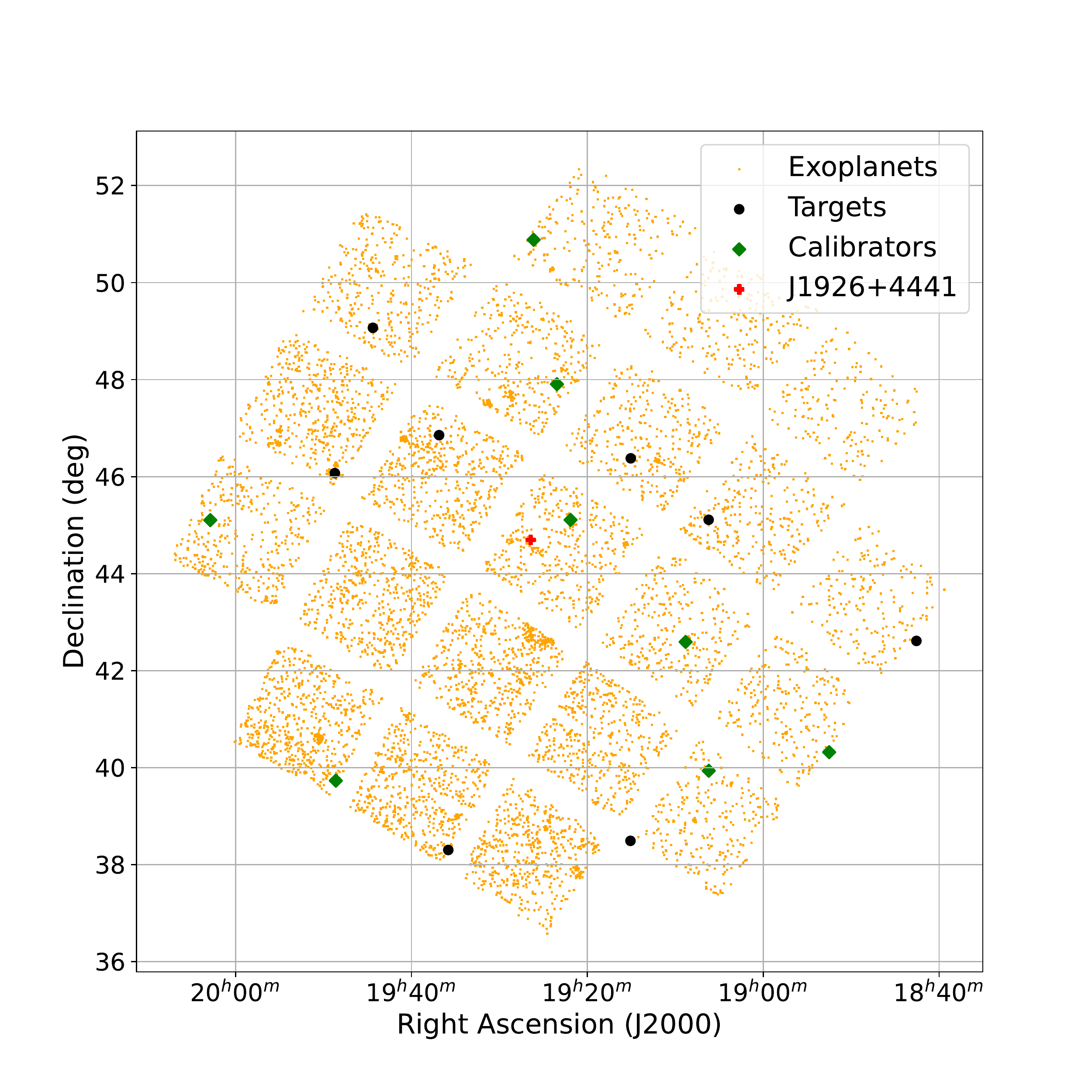}
    \caption{The location of VLBA primary calibrators (green dots) and the secondary targets (blue dots) overplotted on exoplanet detections in the Kepler field. J1926+4441 is shown in red.}
    \label{fig:kepler_field}
\end{figure}

There are nine primary VLBA calibrators located within the Kepler field, including J1926+4441. Unfortunately, the VLA FIRST survey \citep{FIRST1994} does not cover the area of sky of the Kepler field and  we were therefore forced to select our secondary calibrator targets from the lower resolution VLA NVSS \citep{Condon}. Our sample was selected on the basis of angular size ($\theta<20\arcsec$) and flux density ($S > 40\ \rm{mJy}$). Unfortunately, we had no spectral index information about the sources. Eight sources were identified that matched our constraints (see Fig.~\ref{fig:kepler_field}) and Table~\ref{table:coodinates_rsg12}.

\begin{table*}
	\caption{Name, coordinates and flux densities of the secondary targets drawn from NVSS and the primary VLBA calibrators used to phase calibrate the data. These sources were observed in experiment RSG12.}
    \begin{tabular}{lllll} 
		\hline
           
		Name & RA (J2000) & Dec (J2000) & NVSS flux density ($\rm{mJy}$) & VLBA calibrator (ICRF) \\
		\hline
        NVSS 191503+462235 & 19 15 03.77 & +46 22 35.70 & 48.6$\pm$1.5 & J1921+4506\\ 
        NVSS 190612+450634 & 19 06 12.47 & +45 06 34.80 & 815.2$\pm$28.7 & J1908+4235\\
        NVSS 193651+465115 &19 36 51.07 & +46 51 15.60 & 239.4$\pm$8.5 & J1923+4754 \\ 
        NVSS 194842+460415	& 19 48 42.56 & +46 04 15.50 & 100.5$\pm$3.0 & J2002+4506 \\
        NVSS 194422+490401	& 19 44 22.69 & +49 04 01.30 & 40.7$\pm$1.3 & J1926+5052\\ 
        NVSS 191505+382926 & 19 15 05.72 & +38 29 26.80 & 83.4$\pm$2.5 & J1906+3956 \\
        NVSS 193548+381809 & 19 35 48.52 & +38 18 09.10 & 116.5$\pm$3.5 & J1948+3943\\
        NVSS 184234+423645	& 18 42 34.99 & +42 36 45.60 & 157.0$\pm$4.7& J1852+4019 \\
        NVSS 192626+444149	& 19 26 26.57 & +44 41 49.23 & 31.1$\pm$1 & N/A \\
		\hline
	\end{tabular}
 \label{table:coodinates_rsg12}
\end{table*}

\subsection{Observations}

\subsubsection{EVN observations (RSG12)}

Observations (experiment code RSG12) were made on 6 November 2019, 1000-1300 UTC, using the EVN at L-band, with the middle of the observing band centred on the \ion{H}{i} line at 1420.4 MHz. A single 8 MHz IF was recorded in both right and left-hand circular polarization. The 8 MHz band was chosen as a compromise between sensitivity and  spectral resolution. The latter was required in order to further develop the VLBI SETI technique (see section~\ref{section:seti}). 

We employed the technique of phase referencing with a cycle time of 2+7 minutes, cycling between the primary calibrator and secondary candidates. Two of these scans were made for each primary/secondary calibrator pair. Two 10-minute scans were made on J1926+4441. A single scan of 5 minutes was made on the fringe-finder J1800+3848.

Eleven antennas, Jodrell Bank Lovell (Jb1), Westerbork (Wb), Effelsberg (Ef), Medicina (Mc), Noto (Nt), Onsala85 (O8), Torun (Tr), Svetloe (Sv), Zelenchukskayaa (Zc), Badary (Bd), Sardinia (Sr) participated in the observation.

\begin{table}
	\caption{Name and coordinates for sources observed in experiment RSW02}
    \begin{tabular}{lll} 
    \hline
        Name & RA (J2000) & Dec (J2000) \\
    \hline
        3C345 & 16 42 58.80997043 & +39 48 36.9939552 \\
        J1923+4754 & 19 23 27.22979949 & +47 54 16.8181121 \\
        J1926+4441 & 19 26 26.56921500 & +44 41 49.2311000 \\
        J1936+2357 & 19 36 00.92484600 & +23 57 31.9758900 \\ 
    \hline
\end{tabular}
 \label{table:coodinates_rsw02}
\end{table}

 \subsubsection{$e$-MERLIN follow-up observations}

A follow-up observation of the VLBA calibrator J1926+4441 was requested in order to independently analyse a spectral feature first detected in the EVN observations (see section~\ref{section:feature_hi} for more details).  Observations of J1926+4441, together with a phase referencing calibrator J1923+4754, the standard flux calibrator 3C286 (1331+3030), a bandpass calibrator OQ208 (1407+2827), and a bright calibrator 3C84 (0319+4130) were carried out on 11 January 2022, 1100-2100 UTC. The phase referencing cycle time was 10+2 minutes. The observations involved a subset of the $e$-MERLIN array: Jodrell Bank Mark2 (Mk2), Cambridge (Cm), Pickmere (Pi) Defford (De) and Darnhall (Da). The telescope at Knockin (Kn) failed during our observations. 

\subsubsection{EVN follow-up observations (RSW02)}

A follow-up observation with the EVN (experiment code RSW02) was made on 3 March 2022,1610-1700 UTC. As in the case of the e-MERLIN observations, these were requested in order to further probe the nature of an expected spectral feature detected in the original EVN observations (see again section ~\ref{section:feature_hi} for more details). A frequency setup similar to that of the earlier experiment was used but this time with two overlapping 8 MHz bands, the first band at a central reference frequency of 1420.4 MHz and the second offset by 1 MHz with a central reference frequency of 1421.4 MHz. Two bands were chosen to ensure that the feature was not an artefact introduced by correlating a single band. The tuning offset of 1MHz also ensured that the feature in the second band would also be relatively centred.

Four sources (see Table~\ref{table:coodinates_rsw02}) were observed; 3C345 ($\rm{b}\sim40.95\degr$), a fringe finder located at high galactic latitude, J1936+2357 ($\rm{b}\sim1.60\degr$), a source lying close to the galactic plane and J1926+4441 observed previously in RSG12, located within the field. A single scan of 3C345 was made at the beginning of the experiment for 5 minutes with J1926+4441 phase referenced with respect to VLBA calibrator J1923+4752 - a cycle time of 10+2 minutes was employed. Two 10+2 minutes scans were made for J1926+4441/J1923+4752. Finally, a single 10 minute scan of J1936+2357 was made.

Thirteen EVN antennas were originally scheduled for experiment RSW02 but due to various failures only 7 antennas produced useful data: Jb1, Wb, Nt, O8, T6, Ur, Tr, and Ir. 

\subsubsection{EVN data correlation}

The data were correlated using the EVN software correlator at JIVE \citep[SFXC; ][]{Keimpema2015}. The data associated with experiment RSG12 were correlated in two passes. The first and main pass, was applied to all the sources and used 32 frequency channels and 2 seconds of integration time, giving a channel width of 250 kHz. A second high-resolution correlator pass was made for the scans of J1926+4441. This generated 16384 frequency channels each with a width of 488~Hz. An integration time of 0.25 seconds was employed. The output was pipelined into FITS-IDI in two passes with the high-resolution correlation pass of J1926+4441 yielding about $\sim 108\ \rm{GB}$ of data.

Besides the advantage of correlating the data with fine spectral resolution as required by narrow-band SETI searches, this approach also preserves a very large field of view. The limit on the field of view is set by the integration time of 0.25 seconds - time smearing limits the radial field of view to $\sim 4 \arcmin$ for a loss of 10\% at the edge of the field. We estimate the theoretical $1\sigma$ noise level in a naturally weighted imaged to be $0.13\ \rm{mJy}$ for the targets observed for 14 minutes, $0.23\ \rm{mJy}$ for the VLBA phase referencing calibrators observed for 4 minutes and $0.11\ \rm{mJy}$ for J1926+4441 which was observed for 20 minutes.

All the data from EVN experiment RSW02 were correlated with 16384 frequency channels matching the channel size of the previous observation. However, the integration time was increased from 0.25 to 2 seconds to keep the data rates manageable.

\subsubsection{$e$-MERLIN Data correlation}

The $e$-MERLIN data were also correlated in two passes; the first was a continuum pass made over 512 MHz, correlated with 4096 frequency channels each of width 0.125 MHz. The second pass involved the correlation of three "zoom bands" each 0.25 MHz wide, resulting in a combined bandwidth of 0.75 MHz and centred on 1420.4 MHz. Each of the zoom bands was correlated in all 4 Stokes parameters using 512 frequency channels and 4 seconds integration time, resulting in channel sizes similar to those of the high-resolution EVN data.

\subsection{Data analysis}

\subsubsection{EVN experiment RSG12}
The VLBI data analysis was performed in CASA, the Common Astronomy Software Applications \citep{CASA,CASAVLBI}. Custom scripts from JIVE were used to attach the telescope system temperatures and gain curves to the data.  The rudimentary apriori flagging file generated by JIVE was also converted into a format that can be read in CASA. We created two measurement sets, a low spectral resolution data set associated with the 8 primary and secondary calibrators and a high spectral resolution data set associated with J1926+4441.

We inspected the system temperatures and were required to edit and correct some of the entries, in particular for the Zelenchukskaya and Westerbork antennas. The tables were then used to amplitude calibrate the data. We also flagged the outer and inner channels of the 8 MHz band due to edge effects. For the low-resolution data, we flagged $\sim3$ channels at each edge of the band and for the high-resolution data, we flagged $\sim500$ channels. A custom flagging file was created to flag known terrestrial RFI. The auto-correlations for the low-resolution data were flagged while those for the high-resolution data were first plotted (see section~\ref{section:feature_hi}) and then flagged for imaging purposes.

\noindent
For the low-resolution data (including the J1926+4441 data from the first pass), J1926+5052 was used to derive the bandpass calibrations and the solutions (SNR>4) were applied to all sources. The primary phase calibrators were fringe-fitted using the CASA task \textit{fringefit}, and the solutions ($\rm{SNR} > 7$) were transferred to the secondary calibrators. Naturally, J1926+4441 was bright enough to fringe fit on itself.

Since the synthesized beam of NVSS is quite large ($\theta \sim45\arcsec$) it was necessary to make maps that were big enough to encompass possible errors in the NVSS source position. According to \citet{Condon} the positional uncertainty errors $\sigma_p$ are given by:  

\begin{equation}
    \sigma_p = \left(\frac{\sigma \theta}{2 S_p}\right),
    \label{eqn:positional_uncertainty}
\end{equation}

where $\theta$ is the synthesized beam, $S_p$ is the peak intensity in the map and $\sigma$ is the r.m.s noise. In the case of NVSS with a synthesised beam of $45\arcsec$, assuming a detection cutoff of $5\sigma$, the positional uncertainty would be expected to be $\sim4.5\arcsec$ whereas the VLBA calibrators obtained from the  Astrogeo VLBI Calibrator database \footnote{\url{http://astrogeo.org/}} have positional uncertainty errors $<1 
 \textrm{mas}$. We, therefore, used the CASA task \textit{tclean} to make large dirty maps of size $10240\times10240$ pixels ($10 \times 10 \arcsec$), locating the position of peak intensity, making a $256\times256$ pixel map and searching for emission with a $\rm{SNR} > 7$. No obvious detections were noted. 

In the case of J1926+4441, we produced a clean map from the low-resolution data and then applied one round of phase-only self-calibration using an interval of 3 minutes, and then a further round of amplitude self-calibration at an interval of 5 minutes. All the calibration tables from this source were then copied to the high-resolution data set using the CASA task \textit{applycal}. After the calibration tables are copied, small maps were made using the high-resolution data to ensure the calibration tables had been copied and correctly applied. The response of J1926+4441 was subtracted from the high-resolution data set using the CASA task \textit{uvcontsub}.  The data were then phase rotated on-the-fly using the \textit{phasecenter} parameter during imaging in \textit{tclean} to the location of Kepler-111~b, taking into account its position and proper motion as determined by Gaia \citep{Gaia2016,Gaia2018}.  

\subsubsection{e-MERLIN data}
The $e$-MERLIN data were not fully analysed, as they were mainly used as a sanity check on the spectral feature observed in the original EVN data, as discussed in the next section. The $e$-MERLIN data had to be reprocessed to include the auto-correlations that were flagged by the automatic calibration pipeline employed by the observatory.

\subsubsection{EVN experiment RSW02}

The data associated with EVN experiment RSW02 were not calibrated or further processed because we were only interested in the raw auto and cross-correlation data.

\section{Results and Discussion}

\subsection{NVSS targets and J1926+4441}

For EVN experiment RSG12, no detections with a signal-to-noise ratio greater than five were obtained for the secondary calibrator targets. A similar result was noted from the independent and automatic JIVE pipeline analysis using {\sc AIPS}.  Previous snapshot surveys of this type, have usually faired better in terms of detection statistics but these have often preferentially selected flat spectrum objects or used higher resolution finding surveys \citep[e.g.][]{Garrington1999,Beasely2002,Mosoni2004,Wrobel2004}. We note that all these targets have now been detected  by the Very Large Array Sky Survey (VLASS) \citep{VLASS2020} in the quicklook images \citep{QuicklookEpoch12020}. Our source selection was made before VLASS data were publicly available and it turns out that our targets all have very steep spectral indices in the range $\left(S_v\propto\nu^{\alpha}\right)$ $\alpha = -1.67\ \rm{to}\ -2.25$. The true spectral indices are probably much less than this but resolution effects are at work here - in particular, the VLASS restoring beam (2.5\arcsec) is more than two orders of magnitude larger in beam area than the EVN restoring beam.  It is therefore probably not very surprising that we do not detect any of these sources on VLBI scales.

\begin{figure*}
  \centering
  \begin{tabular}{cc}

      \includegraphics[width=.9\columnwidth]{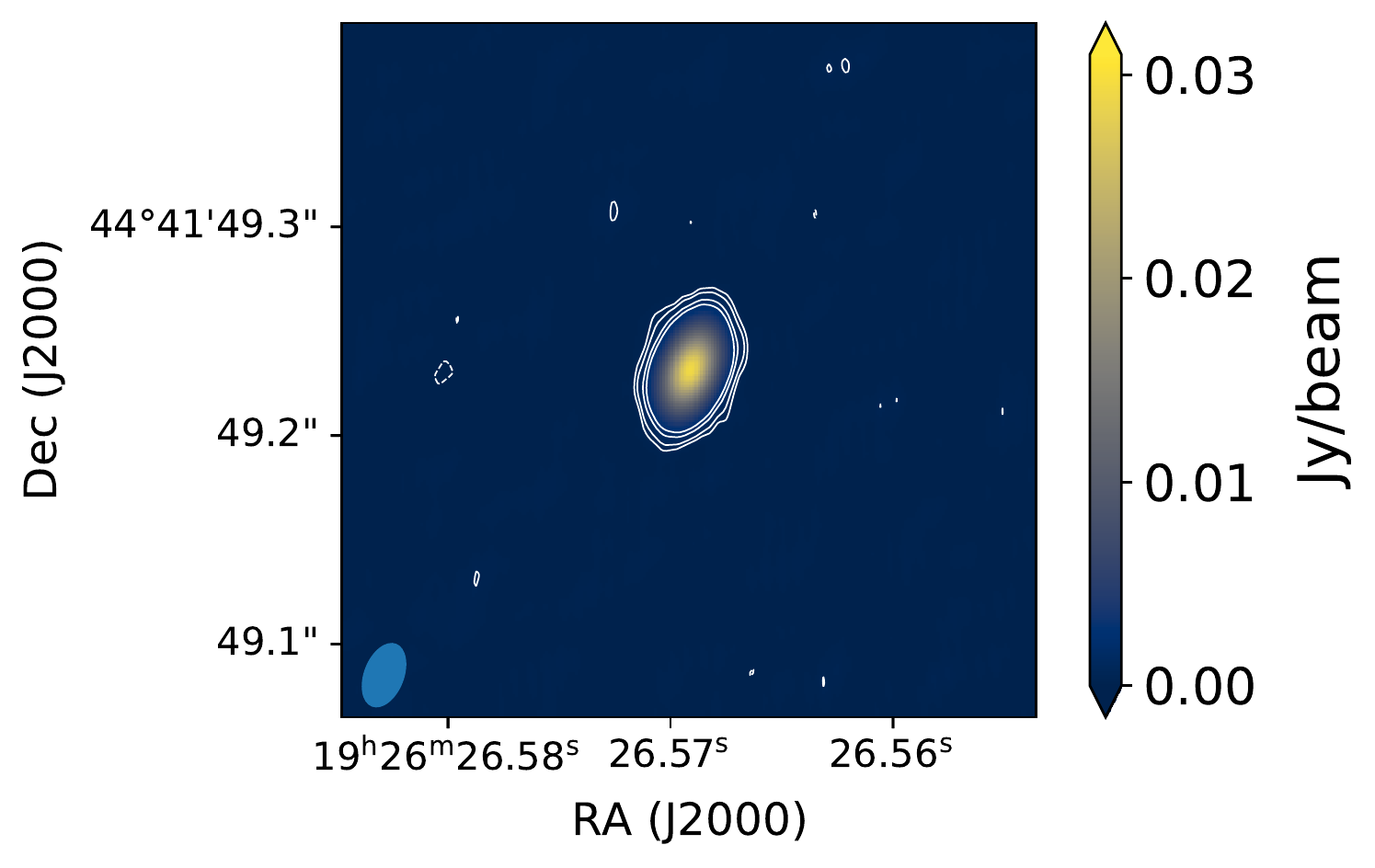} &
      \includegraphics[width=.9\columnwidth]{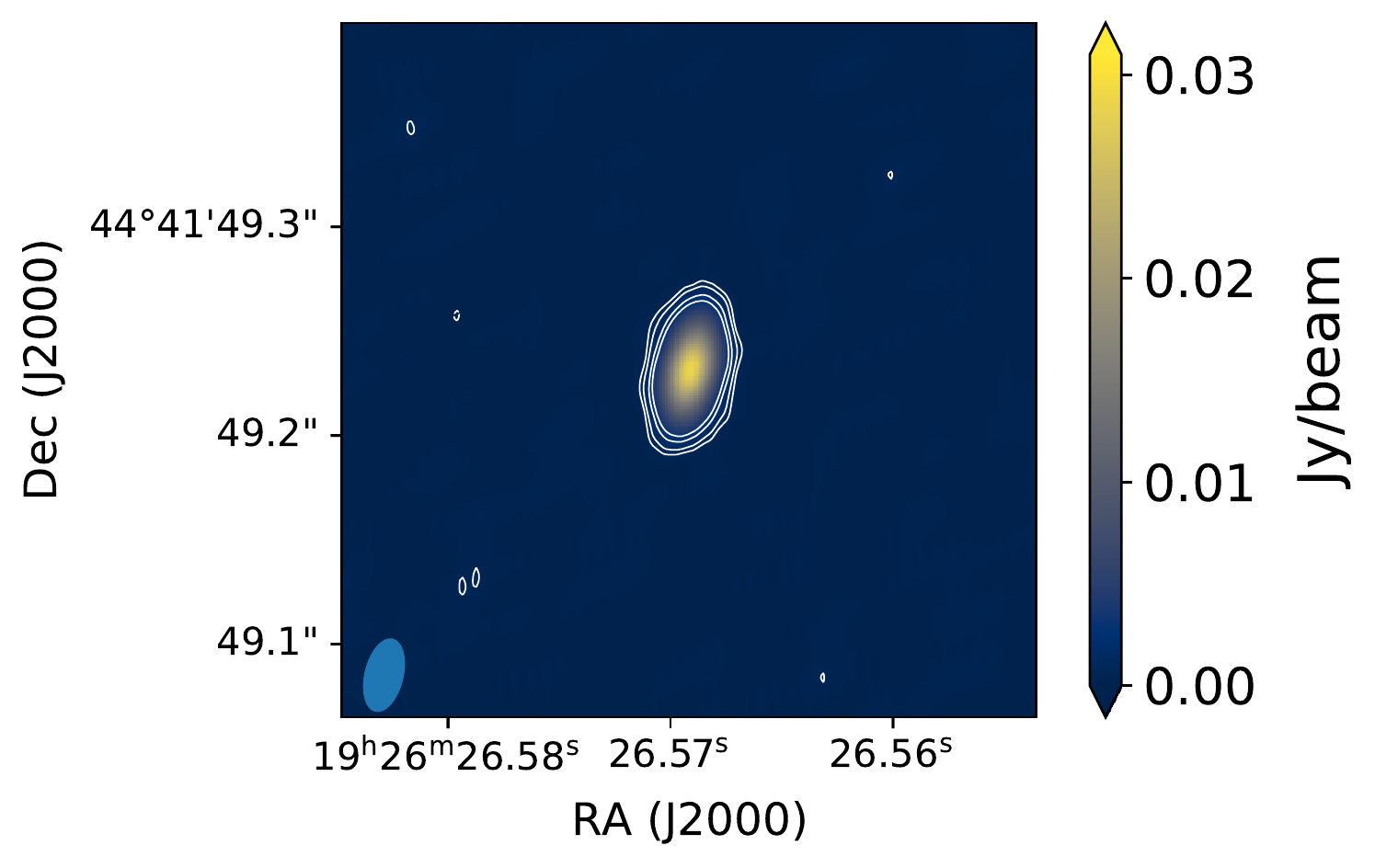} \\
      
      (a) & (b)
  \end{tabular}
 \caption{(a)The final CLEAN map has a peak brightness of $\sim29.3$ mJy/beam and an r.m.s noise of $\sim0.14$ mJy/beam, giving a signal-to-ratio of $\sim209$.  (b) The final CLEAN map after applying the phase and amplitude corrections from the low-resolution data set. The map has peak brightness of $\sim28.9$ mJy/beam and the r.m.s. noise of $\sim0.2$ mJy/beam resulting in a signal-to-noise ratio of $\sim$145. The maps are presented with  contours drawn at intervals of (-3,3,5,10,15)$\sigma$. The colour bar is shown to the right of the respective images in units of Jy/beam. The light blue circle in both maps represents the restoring beam.}

  \label{fig:cont}
\end{figure*}

Naturally weighted dirty and CLEANed images of J1926+4441 were generated from the low spectral resolution data - see Fig.~\ref{fig:cont}. The images presented in Fig.~\ref{fig:cont}(b) were generated from the high-resolution data, after applying the calibration tables from the low spectral resolution data. This appears to have worked well, and greatly reduces the computational burden of calibrating the large, high-resolution data set.

\subsection{Feature in the data}
\label{section:feature_hi}

\begin{figure}
    \centering
    \includegraphics[width=.8\columnwidth]{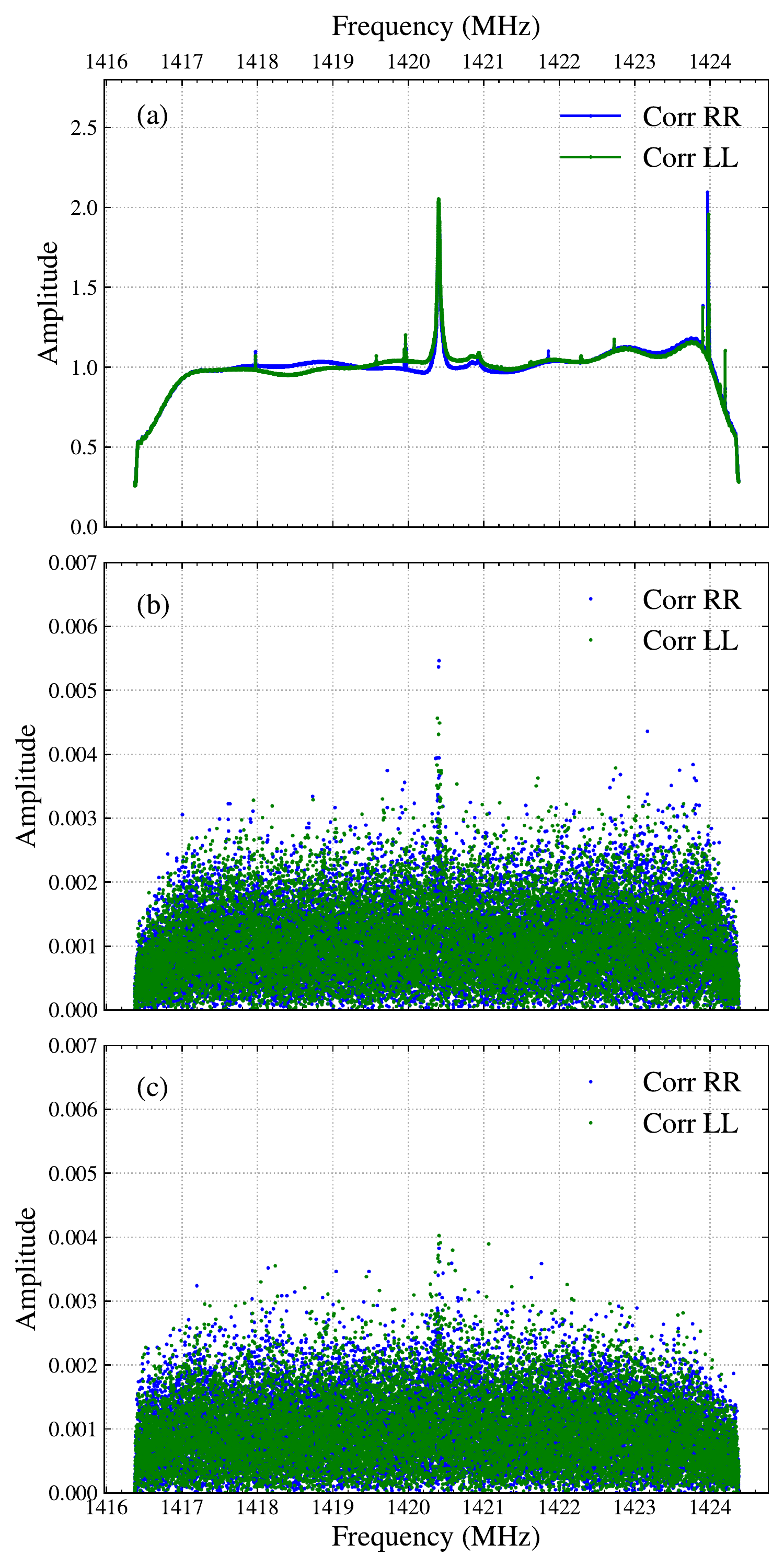}
    \caption{(a) The auto-correlations of antenna Effelsberg showing the right and left circular polarization.
    (b) The cross-correlations formed by the shortest baseline ($\sim267\ \rm{km}$) to the most sensitive antenna, Effelsberg paired with the antenna Westerbork . The right and left circular polarization hands are shown in the colours blue and green respectively. 
    (c) The cross-correlations formed by the longest baseline ($\sim5930 \rm{km}$) to the most sensitive antenna, Effelsberg paired with the antenna Badary. The right and left circular polarization hands are shown in the colours blue and green respectively. All the amplitudes are in arbitrary correlator units and the data are plotted over the entire 8 MHz band ranging from 1416.4 to 1424.4 MHz.}
    \label{fig:J1926_autos_cross}
\end{figure}

In the course of visually inspecting the high-resolution data, we noticed  a feature in the auto-correlations shown in Fig.~\ref{fig:J1926_autos_cross}(a) that peaks at $\sim 1420.424\pm0.002\ \rm{MHz}$ and is about $\sim0.25\ \rm{MHz}$ in width from $\sim1420.3$ to $\sim1420.55\ \rm{MHz}$. It is not uncommon to see such features in VLBI auto-correlations due to the presence of foreground, galactic neutral hydrogen (\ion{H}{i}) and the field with galactic coordinates $l=76.35^\circ, b=+13.5^\circ$ lies close to the galactic plane (see Fig.~\ref{fig:kepler_field}). 

What was more surprising is that in the high-resolution data we also see the same feature appearing in the cross-correlations, as shown in Fig.~\ref{fig:J1926_autos_cross}(b) and Fig.~\ref{fig:J1926_autos_cross}(c). This feature is very comparable in its characteristics to that seen in the auto-correlations - indeed it is located in the same spectral channel. The feature is $\sim0.25\ \rm{MHz}$ in width, spanning the same frequency channels (1420.3 to $\sim$ 1420.55 MHz) as in the auto-correlations. 

\begin{figure}
    \centering
    \includegraphics[width=.9\columnwidth]{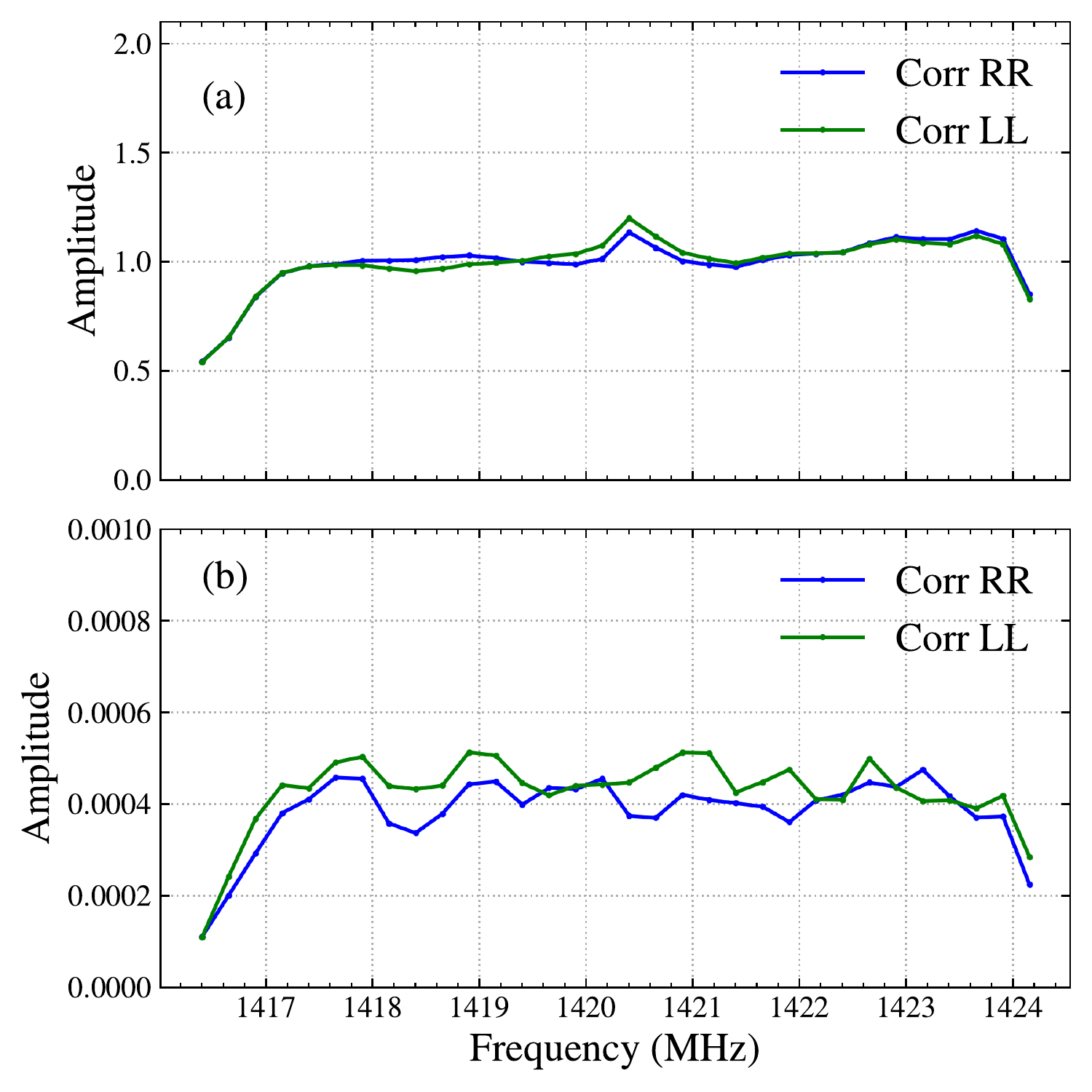}
    \caption{(a) The auto-correlations of antenna Effelsberg from the low-resolution data showing the right and left circularly polarized hands. The spectral feature can still be identified in the plot. (b) The cross-correlations formed by antennas Effelsberg and Westerbork show the two circularly polarized hands. The spectral feature is no longer visible in this plot. All the amplitudes are in arbitrary correlator units and the data are plotted over the entire 8 MHz band ranging from 1416.4 to 1424.4 MHz.}
    \label{fig:J1926_low_res}
\end{figure}

We also see these features in the auto-correlations of the low-resolution data (see Fig.~\ref{fig:J1926_low_res}(a)) but they are much more difficult to recognise, and are no longer visible in the cross-correlation data at all (see Fig.~\ref{fig:J1926_low_res}(b)). The spectral feature in the low resolution auto-correlation data would almost certainly have gone unnoticed if we had not been prompted to look for it via the earlier discovery of this feature in the high-resolution data.

\begin{figure}
    \centering
    \includegraphics[width=.8\columnwidth]{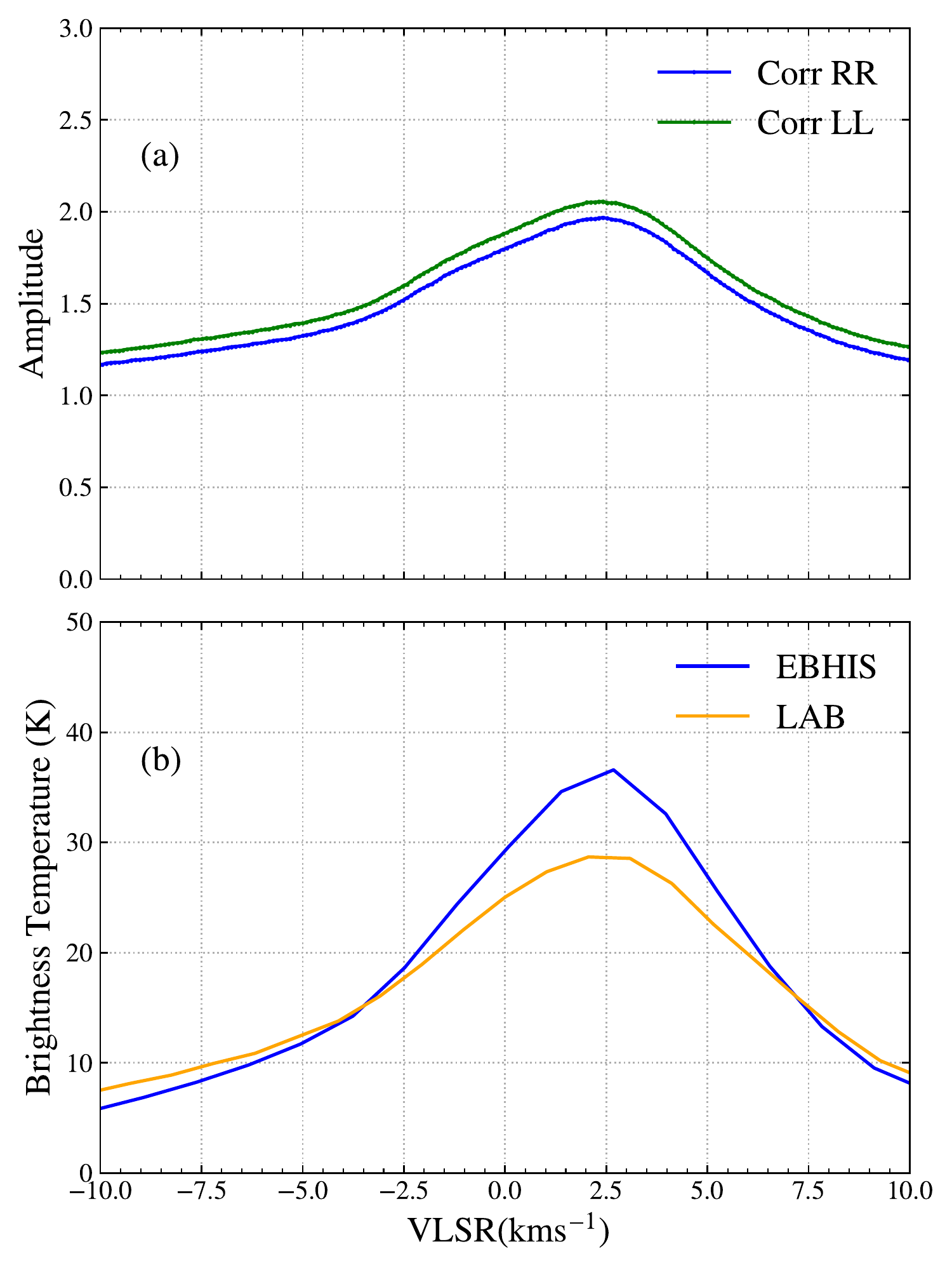}
    \caption{(a) The LSRK velocity for the observed feature in the auto-correlations of antenna Effelsberg showing the right and the left circularly polarized hands  (b) The velocity profile of J1926+4441 obtained from the LAB and EHBIS galactic hydrogen surveys. The data are plotted in the range of -10 \kms to 10 \kms.}
    \label{fig:velocity_evn_data}
\end{figure}

To test whether the spectral feature was associated with galactic \ion{H}{i}, we performed local standard of rest corrections to the EVN auto-correlation data using the CASA task cvel and compared this with the LAB \citep{LAB} and EBHIS \citep{EBHIS} \ion{H}{i} surveys \footnote{\raggedright{\url{https://www.astro.uni-bonn.de/hisurvey/AllSky_profiles/index.php}}}. For the EVN data, the VLSR velocity of the feature was found to peak at $2.4\pm0.1$\kms. After querying the EBHIS \citep{EBHIS} and the LAB \citep{LAB} using the default beamwidth size of 0.2\degr, the \ion{H}{i} profile shows that the velocity they measure peaks at 2.06\kms and 2.68\kms respectively, strongly suggesting that the feature we see in the EVN auto-correlations is due, as we suspected, to galactic \ion{H}{i} in emission. Fig.\ref{fig:velocity_evn_data} compares the profile we observe with the profile observed by LAB and EBHIS.

\begin{figure}
    \centering
    \includegraphics[width=.8\columnwidth]{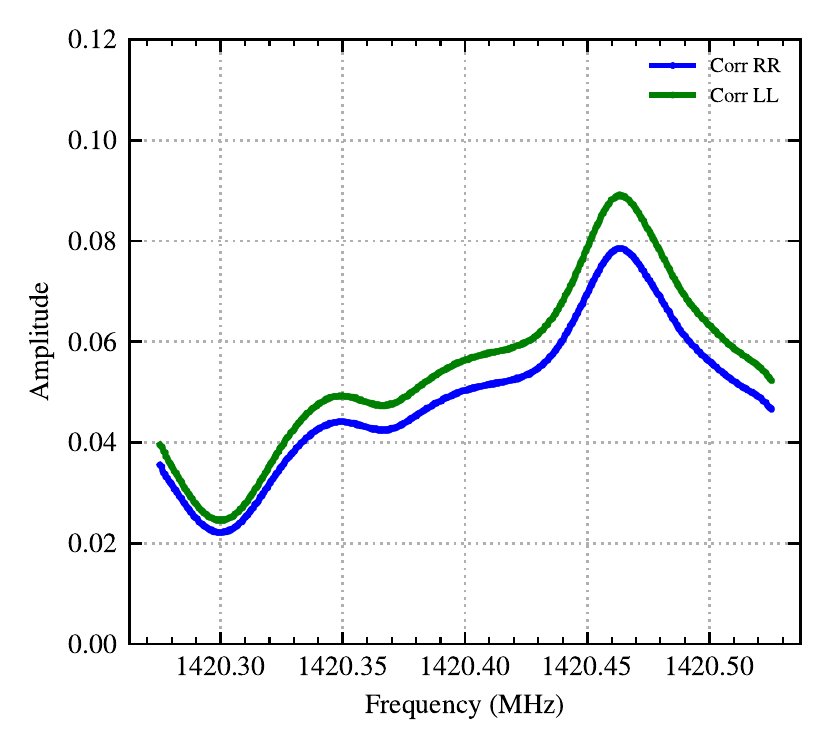}
    \caption{The auto-correlations of antenna Mark 2 showing the right and left circularly polarization hands in a 0.25 MHz spectral window centred on 1420.4 MHz. The amplitude is in arbitrary correlator units.}
    \label{fig:emerlin_mk2}
\end{figure}

\begin{figure}
    \centering
    \includegraphics[width=.8\columnwidth]{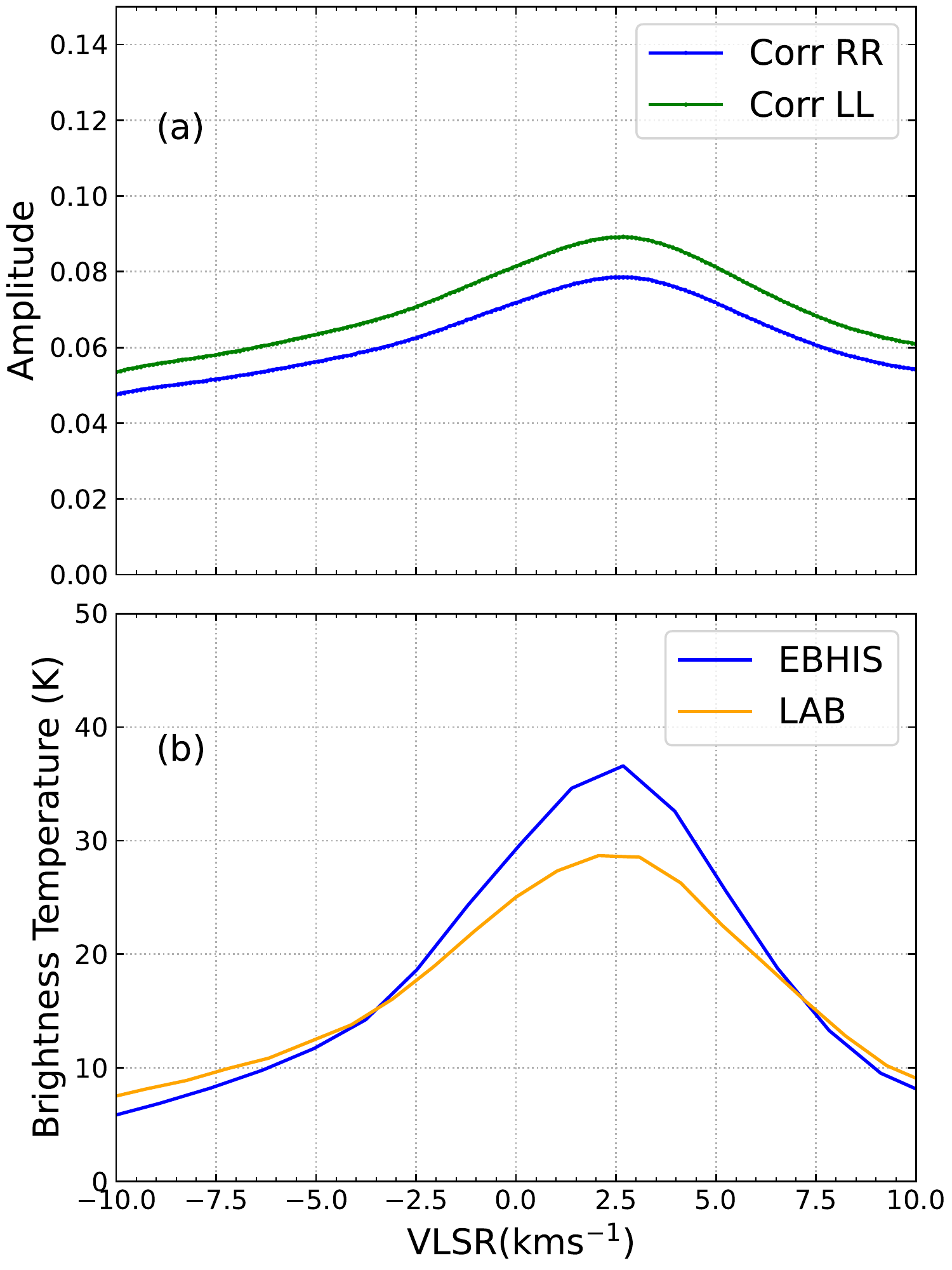}
    \caption{(a) The LSRK velocity for the observed feature in the auto-correlations of antenna Mark 2 showing the right and the left circularly polarized hands  (b) The velocity profile of J1926+4441 obtained from the LAB and EHBIS galactic hydrogen surveys. The data are plotted in the range of -10 \kms to 10 \kms.}
    \label{fig:emerlin_mk2_velocity}
\end{figure}

The $e$-MERLIN auto-correlations of J1926+4441 also show the same feature peaking at a frequency of about $\sim1420.462\pm0.002 \ \rm{MHz}$ in the uncorrected geocentric frame (see Fig.~\ref{fig:emerlin_mk2}). Application of LSR corrections and subsequent comparison with velocity profiles from \citep[][]{EBHIS,LAB}, show a similar profile and a peak velocity at a velocity of $2.7\pm0.1$\kms (see Fig.~\ref{fig:emerlin_mk2_velocity}) conclusively proving that the feature in the auto-correlations is galactic \ion{H}{i}. However, the feature is not detected in the $e$-MERLIN cross-correlation data.

\subsubsection{Nature of the feature} 

The comparison of the velocities measured here and in EBHIS \citep{EBHIS} and LAB \citep{LAB}, clearly demonstrate that the feature in both the EVN and $e$-MERLIN auto-correlations is galactic \ion{H}{i} in emission. Since \ion{H}{i} in emission has a brightness temperature $< 10^4\ \rm{K}$ \citep{Roy2013}, it is not plausible that it could be detected in the EVN data - for example, even the shortest and most sensitive EVN baseline has a brightness sensitivity $\sim 10^5\ \rm{K}$. 

As a further test, we attempted to make EVN images of this feature by making maps that only included those channels with significant emission (channels 8100 to 8400). No detections of the feature were made in the image plane, also employing very large 10k by 10k images. We also note that the feature persists after the continuum subtraction of J1926+4441 (see section \ref{position}). 

If the feature in the cross-correlations is not a detection of \ion{H}{i} in emission, then what can it be? We believe that this is an artefact of the EVN data that is related to the large increase in telescope receiver temperature due to the presence of bright 21 cm emission in the primary beam. Indeed a similar phenomenon was observed by \citet{Dickey1979} in some early VLA \ion{H}{i} observations. The data from EBHIS and LAB surveys \citep{EBHIS,LAB} imply that the peak \ion{H}{i} emission in this direction on the sky is $\sim36.6\ \rm{K}$. This is comparable to the noise temperatures of the EVN receivers on cold sky. 

\begin{figure}
    \centering
    \includegraphics[width=.8\columnwidth]{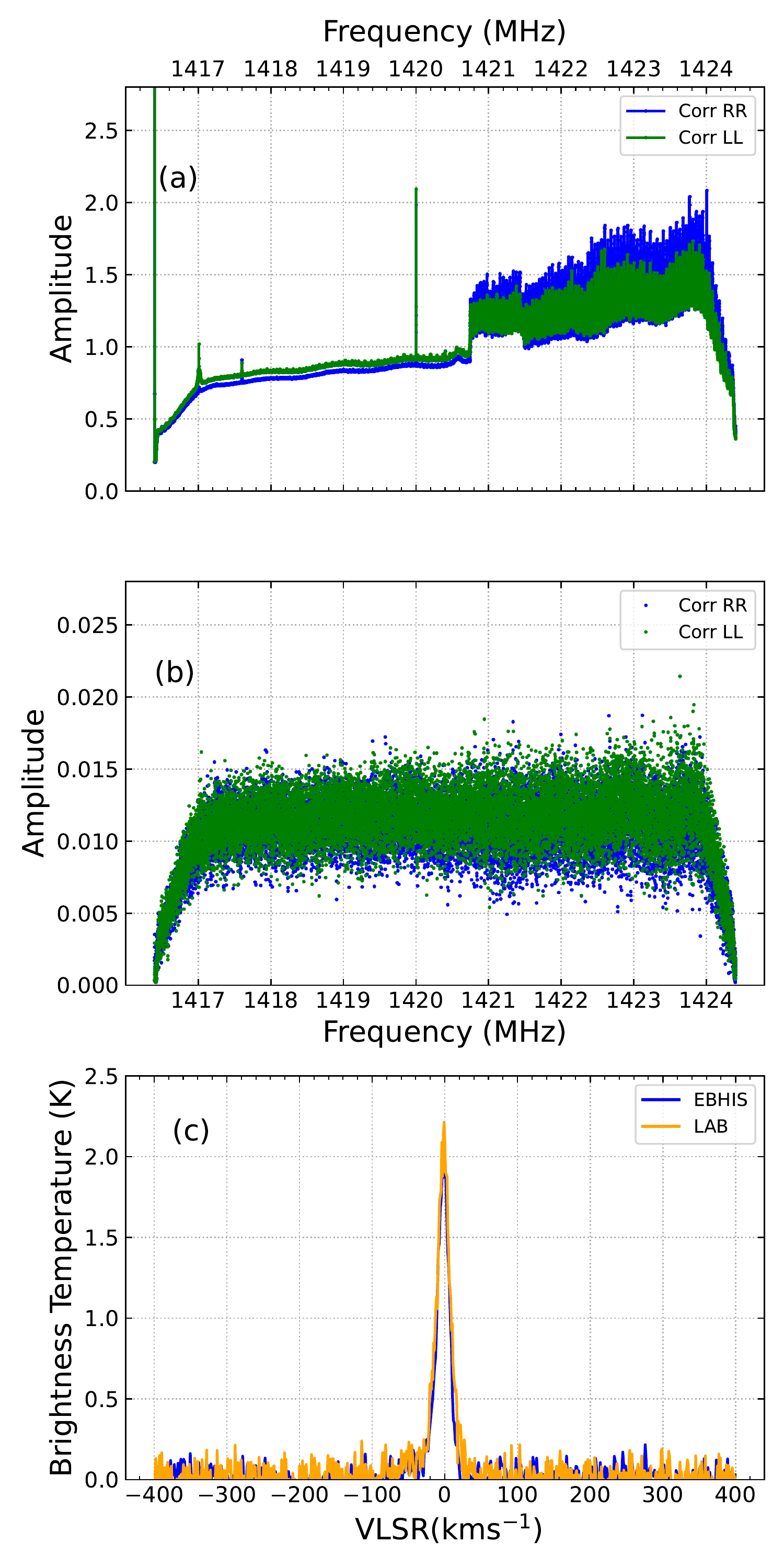}
    \caption{
    (a) The auto-correlations of antenna Westerbork showing the right and left circular polarization formed from observations of 3C345.
    (b) The cross-correlations of the baseline formed by antennas Westerbork and Jodrell Bank  ($\sim597\ \rm{km}$) formed from observations of 3C345. The right and left circular polarization hands are shown in the colours blue and green respectively. 
    (c) The velocity profile of 3C345 obtained from the LAB and EHBIS galactic hydrogen surveys. The data are plotted in the range of -400 \kms to 400 \kms.}
    
    \label{fig:3C345_autos}
    
\end{figure}

\begin{figure}
    \centering
    \includegraphics[width=.8\columnwidth]{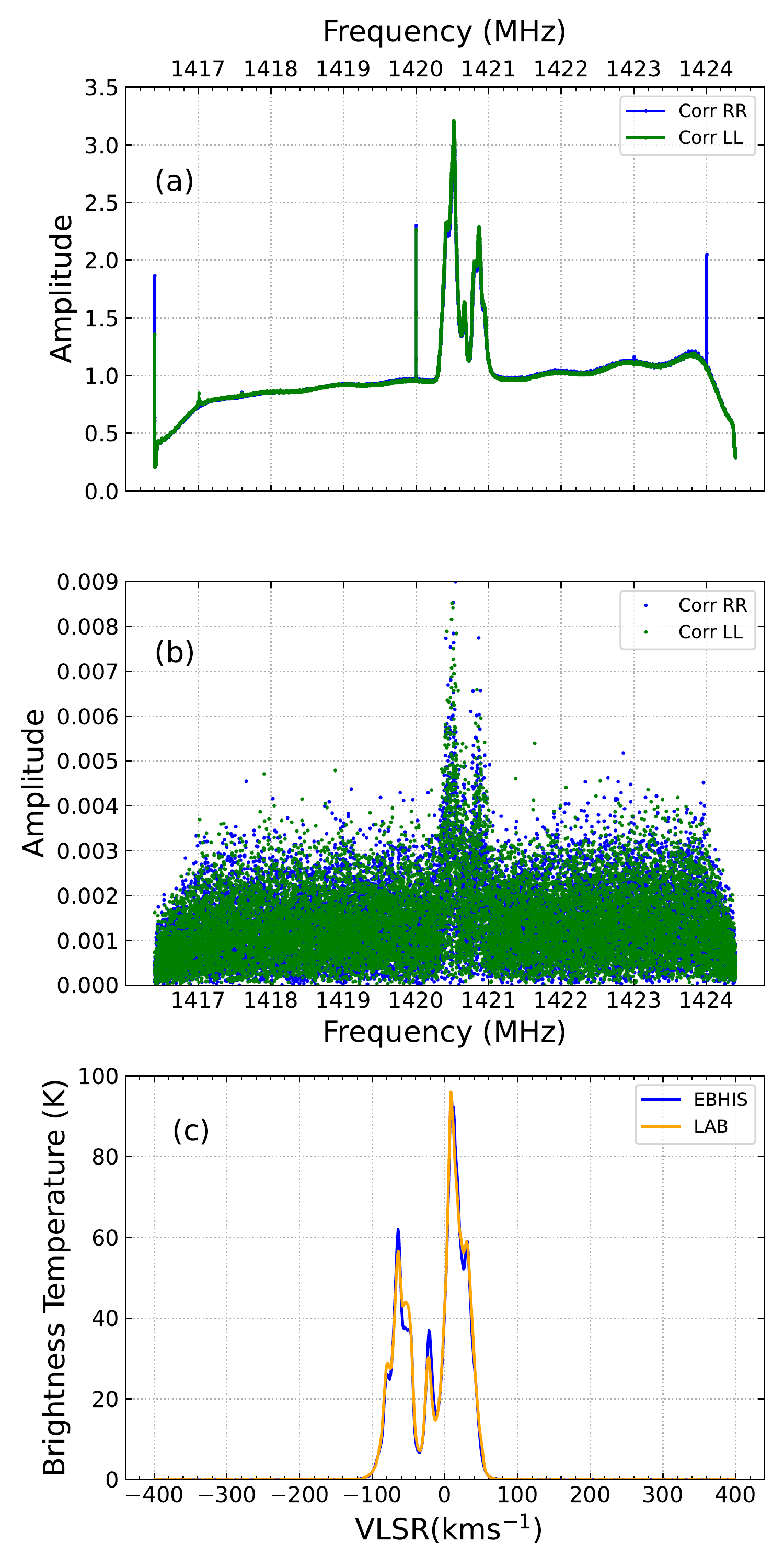}
    \caption{
    (a) The auto-correlations of antenna Westerbork showing the right and left circular polarization formed from observations of 1936+2357 .
    (b) The cross-correlations of the baseline formed by antennas Westerbork and Jodrell Bank  ($\sim597\ \rm{km}$) formed from observations of 1936+2357 . The right and left circular polarization hands are shown in the colours blue and green respectively. 
    (c) The velocity profile of 1936+2357  obtained from the LAB and EHBIS galactic hydrogen surveys. The data are plotted in the range of -400 \kms to 400 \kms.}
    
    \label{fig:J1936_autos}
    
\end{figure}

In order to test this hypothesis, we used the EVN to re-observe J1926+4441 as before but also added in observations of one source at high galactic latitude (3C345) and another at very low galactic latiude (1936+2357). This follow up experiment (RSW02) used the same high-resolution set up as that  employed in the original EVN experiment RSW02. In the case of 3C345, plots of the auto-correlations and cross-correlations (see Fig.~\ref{fig:3C345_autos}(a) and Fig.~\ref{fig:3C345_autos}(b)) show no evidence for any features but for J1936+2357 features are very prominent in both the auto-correlations and cross-correlations - see  Fig.~\ref{fig:J1936_autos}(a) and Fig.~\ref{fig:J1936_autos}(b). The original feature seen in J1926+4441 is re-detected in these observations too.

From these results, we conclude that the unusual feature in the EVN cross-correlation data is indeed associated with the presence of \ion{H}{i} within the telescope beams. The \ion{H}{i} brightness temperature in the region of 3C345 peaks at $2.2$K \citep{LAB,EBHIS} but for J1936+2357 the \ion{H}{i} brightness temperature is measured to be significantly larger $\sim96$K \citep{LAB,EBHIS}. For J1936+2357, the \ion{H}{i} features observed in the auto-correlations and cross-correlations, take the shape of the velocity profile of the \ion{H}{i} gas.

We attempted to simulate the effect of this large increase in receiver noise on the cross-correlation data over a relatively distinct part of our 8 MHz band. The measured visibilities $\vec{Z}$ will consist of the true visibility $\vec{V}$ which will be affected by noise $\vec{\epsilon}$ with independent real and imaginary parts both normally distributed with mean $\mu=0$ and variance $\sigma^2$ \citep{TMS} resulting in a complex normal random variable of the two components.

\begin{figure}
    \centering
    \includegraphics[width=.9\columnwidth]{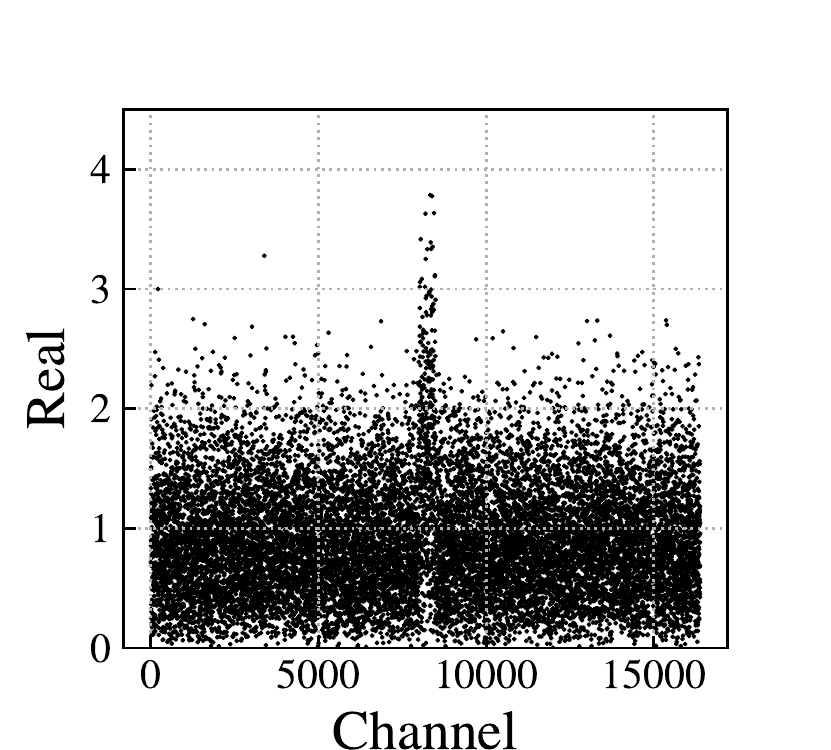}
    \caption{Simulated noise representing the system temperatures with an additional component fitted.}
    \label{fig:noise_sim}
\end{figure}

To simulate this effect, we generated random noise with a variance of 1 over the entire 8 MHz which is divided into 16384 channels, then fitting noise with a variance of 0.75 to channels 8000 to 8500 where the bulk of the feature lies. The resulting plot is shown in Fig.~\ref{fig:noise_sim} which shows the real component of the noise against the channels which looks similar to the feature  seen in the cross-correlations as shown in Fig.~\ref{fig:J1926_autos_cross}(b) and ~\ref{fig:J1926_autos_cross}(c).

We note that the simulated feature is similar to the one present in the EVN data, supporting the hypothesis that the feature is an artefact in the data, associated with the steep rise in the telescopes' system temperatures associated with bright \ion{H}{i} emission filling the primary beams. This artefact is not observed in the $e$-MERLIN cross-correlations. This is likely due to the different normalisation processes employed in the EVN and $e$-MERLIN correlators.  

\subsection{Search for technosignatures in Kepler-111~b}
\label{section:seti}

Most cosmic radio sources exhibit a continuous, broad-band, power-law spectrum since the emission is typically dominated by synchrotron radiation.  While some sources e.g. masers present relatively narrow spectral-line features, these are always broadened by thermal processes. For example, \cite{Cohen1987} showed that the lower limit of the line widths of OH masers is $\sim 500$~Hz. Artificial signals generated by technology are typically much narrower than this, and SETI surveys are thus tuned to detect narrow band signals ($\sim$ {a few Hz}). The relative accelerations between the 
 receiver and the transmitter introduces a Doppler drift $\dot{f}$ in narrow-band signals given as: 

\begin{equation}
    \dot{f} = \frac{d\overline{V}}{dt}\frac{f_{\rm{rest}}}{c} ,
    \label{eqn:drift}
\end{equation}

where $\overline{V}$ is the velocity along the line of sight relative to the transmitter and receiver, $f_{\rm{rest}}$ is the rest frequency at which the signals are transmitted and $c$ is the speed of light.

\subsubsection{Maximum image integration time} 

Assuming a typical maximum drift rate of $2 \ \mathrm{Hz\,s^{-1}}$ \citep[e.g.][]{Enriquez2017} a narrow-band technosignature signal will move from one EVN channel to another in a time $\Delta t_{\rm{int}}$ given by:

\begin{equation}
    \Delta t_{\rm{int}} = \frac{\mathrm{channel \ width}}{\rm{drift \ rate}} = \frac{488\  \rm{Hz}}{2 \ \mathrm{Hz\,s^{-1}}} \sim  240 \ \rm{s}.
\end{equation}

This time scale, therefore, sets a limit on the integration time we can use to produce images of Kepler-111~b - employing longer integration times would wash out the response of Doppler drifting signals with a bandwidth similar to the EVN channel width. For the two ten minutes scans of J1926+4441, we thus divide the data into six 200 seconds chunks resulting in three cubes for each scan, and then made three-dimensional cubes in right ascension, declination and frequency.

\subsubsection{Position of Kepler-111~b}
\label{position}

Kepler-111~b is a super-Earth with a radius $\sim {3.12\rm{R}_\oplus}$ in a tight 3.3 day orbit around a G-type star (Kepler-111).
Located at a distance of $657.223^{+6.014}_{-5.908}$ pc \citep{Bailer-Jones}, we find that the two sources lie within the same EVN image pixel - the maximum angle of separation between the two is indeed only $\phi\sim0.07$ milliarcseconds. We must also take into account the proper motion of the system obtained from GAIA \citep{Gaia2016,Gaia2018}. Using these catalogues we calculated the position of Kepler-111~b to the date of the EVN observations. 

As noted earlier, we subtracted the radio continuum response of J1926+4441 from the high-resolution data and phase rotated the data to the position of Kepler-111. As expected, the spectral feature described earlier (see section \ref{section:feature_hi}) persists - see figure~\ref{fig:J1926_autos_cross}(b) and ~\ref{fig:J1926_autos_cross}(c). We, therefore, flagged the associated channels in our subsequent analysis.

\subsection{SETI signal data analysis} 

\begin{figure*}
    \centering
    \includegraphics[width=\textwidth]{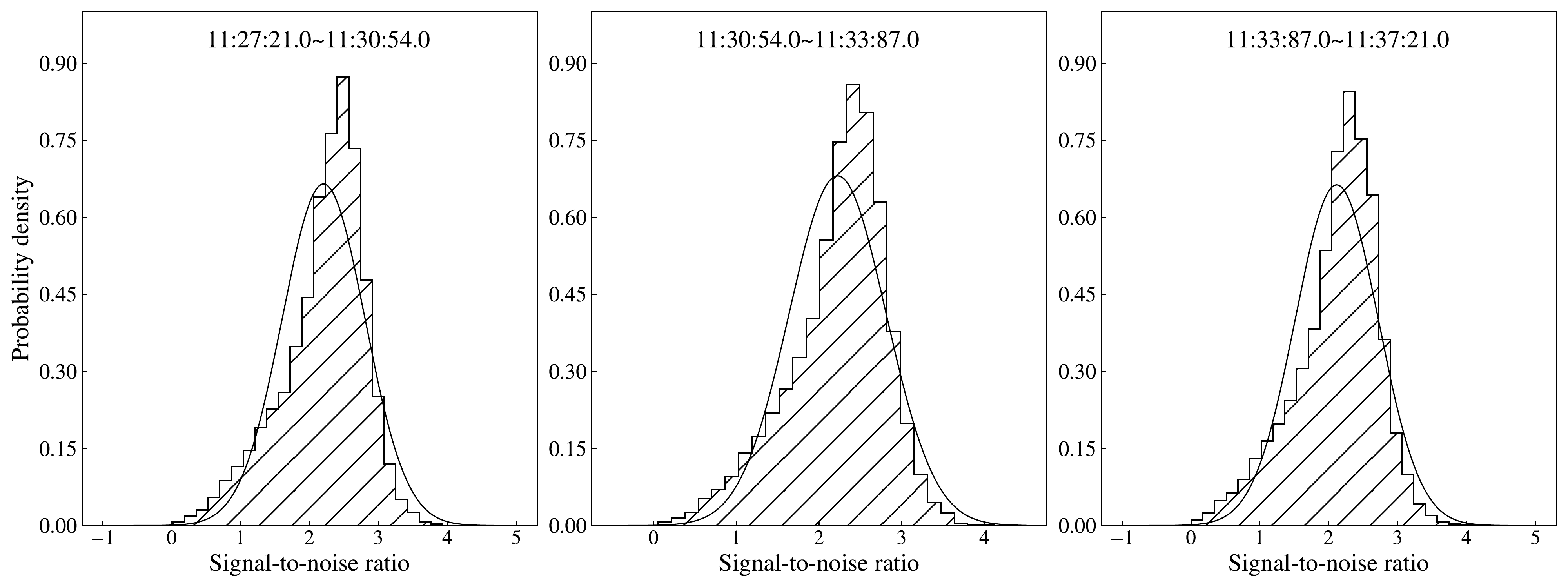}
    \caption{The signal-to-noise ratio of the various cubes made using data from the first scan with Gaussians fitted. The first cube spans the time range 11:27:21.0$\sim$11:30:54.0 and has a mean signal-to-noise ratio of 2.202 and a standard deviation of 0.600, the second cube spans the time range 11:30:54.0$\sim$11:33:87.0 and has a mean signal-to-noise ratio of 2.223 and a standard deviation of 0.586 and the third cube spans the time range 11:33:87.0$\sim$11:37:21.0 and has a mean signal-to-noise ratio of 2.115 and a standard deviation of 0.602. }
    \label{fig:stats_imagecube1}
\end{figure*}

\begin{figure*}
    \centering
    \includegraphics[width=\textwidth]{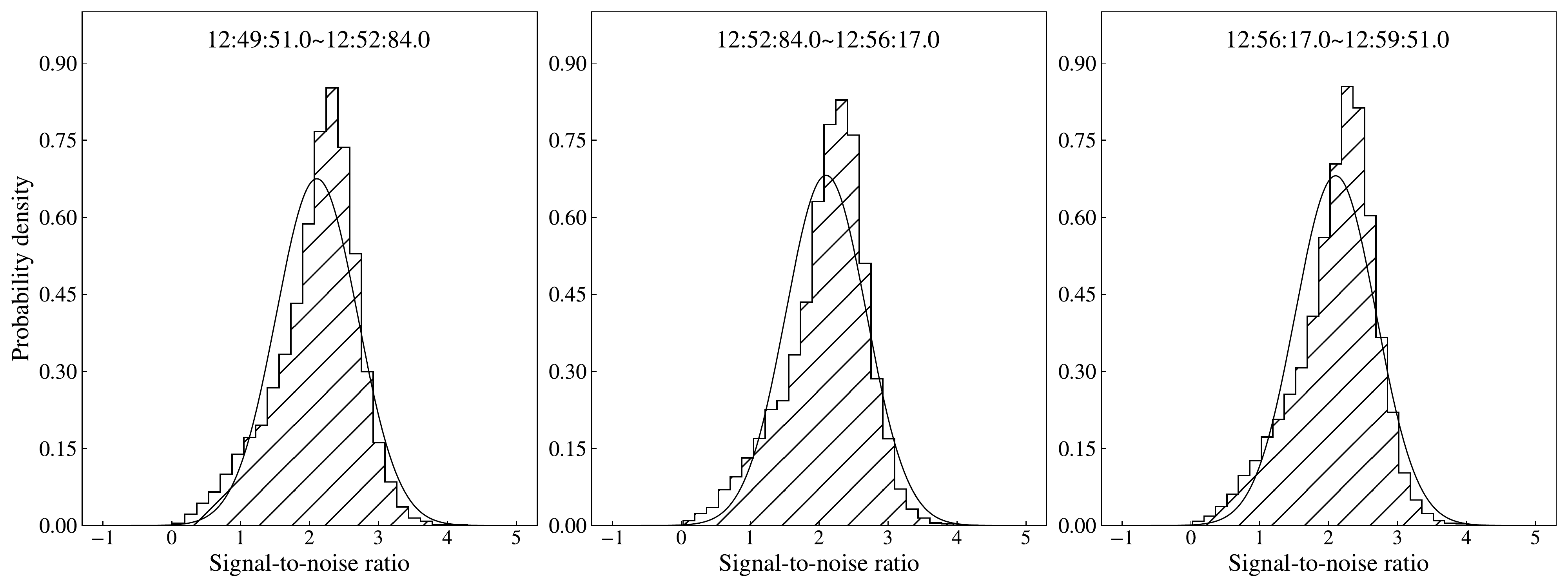}
    \caption{The signal-to-noise ratio of the various cubes made using data from the first scan with Gaussians fitted. The first cube spans the time range 12:49:51.0$\sim$12:52:84.0 and has a mean signal-to-noise ratio of 2.105 and a standard deviation of 0.591, the second cube spans the time range 12:52:84.0$\sim$12:56:17.0 and has a mean signal-to-noise ratio of 2.100 and a standard deviation of 0.585 and the third cube spans the time range 12:56:17.0$\sim$12:59:51.0 and has a mean signal-to-noise ratio of 2.100 and a standard deviation of 0.586. }
    \label{fig:stats_imagecube2}
\end{figure*}

We flagged channels $7900-8600$ (channel 8192 is centered on the \ion{H}{i} line and 1 channel $\sim488\ \rm{Hz}$) from the data set as these were dominated by the spectral feature associated with bright \ion{H}{i}, in addition to $\sim500$ channels at both edges of the 8~MHz band. Image cubes were made using the remaining 14684 frequency channels. Since the location of Kepler-111~b is well known, we were able to make relatively small (32 x 32) images of the field.  For each of the 6 cubes, the peak flux, r.m.s. noise and the signal-to-noise ratio (SNR) were computed for each of the channel maps and the signal-to-noise ratio statistics were analysed by Gaussian fitting (see Fig.~\ref{fig:stats_imagecube1},~\ref{fig:stats_imagecube2}). In this analysis, we do not detect any signals in any channel that have a signal-to-noise ratio in excess of 7.

We calculated the effective isotropic radiated power (EIPR) our observations are sensitive to using: 

\begin{equation}
     \rm{EIRP} = 4\pi \mathit{R^2} \sigma_{\rm{thresh}}\rm{SEFD_{array}}\left(\frac{\delta \nu}{\mathit{n}_{\rm{pol}}\mathit{t}_{int}}\right)^{\frac{1}{2}}.
\end{equation}


For the EVN observations presented here, $\rm{SEFD_{RSG12}}\sim14.4$ Jy and  $\delta_{\nu}$= 488 Hz, $t_{\rm{int}}$ = 240 s and $n_{\rm{pol}}$ = 2. At a distance of $R\sim657$ pc, and assuming a detection as $\sigma_{\rm{thresh}} > 5$, we estimate the EIRP we are sensitive to for Kepler-111~b is $\sim4\times10^{15}$ W. We detect no sources with an EIRP beyond this limit in our data.

VLBI routinely brings together in the same array some of the most sensitive antennas in the world e.g. the Effelsberg and the Lovell with a system equivalent flux density (SEFD) of 20 and 35 Jy at 21 cm respectively. This increases the likelihood of a detection on baselines formed by these antennas. However, comparisons of different SETI experiments are difficult since observing setups and instruments will usually vary, and also integration times. These limitations necessitate the use of a common figure of merit such as the 
continuous wave transmitter figure of merit (CWTFM) defined by \citet{Enriquez2017} as

\begin{equation}
   \rm{CWFTM} = \zeta_{A0}\frac{\rm{EIRP}}{N_{\rm{stars}}\nu_{\rm{frac}}},
   \label{eqn:cwtfm}
\end{equation}

where $N_{\rm{stars}}$ is the number of stars observed with $N$ pointings, $\nu_{\rm{frac}}=\nu_{\rm{tot}}/ \nu_{\rm{mid}}$ is the fractional bandwidth used in the observation and $\zeta_{\rm{A0}}$ is a normalization factor such that CWFTM = 1 when the EIRP is equal to the luminosity of the  Arecibo's planetary radar system $L_{\rm{A0}}=10^{13}$, $N_{\rm{stars}}=1000$ and $\nu_{\rm{frac}}=0.5$. We calculate a CWFTM $\sim10^{7}$. This high value is because of the small fractional bandwidth used for this observation, the targeting of a single star using a single pointing and the relatively large distance to Kepler 111-b (657 pc).

\section{Conclusions}

VLBI offers many advantages for the detection and follow-up of SETI candidates of interest. The advent of software correlators has presented us with the opportunity to exploit these advantages that include less susceptibility to terrestrial RFI added redundancy in the case that a SETI signal does not repeat, and the ability to provide precision locations for SETI transmitters in the image domain. To further develop this technique, we have observed sources selected from the NVSS with the EVN and phase referenced them with known VLBA calibrators in the Kepler field with the intention of identifying secondary phase calibrators that would be useful in future VLBI SETI studies of the same field. None of the potential secondary candidates were detected.  However, EVN observations of one of the main calibration sources, J1926+4441, lying only $1.88\arcmin$ from Kepler-111~b, permitted us to place an upper limit on the prevalence of powerful transmitters in this system. We detect no transmitters with an EIRP $\sim4\times10^{15}$~W at the time of our observations. During a careful analysis of this data set, we noticed a persistent spectral feature in the auto-correlation and cross-correlation data. In both cases, this is associated with galactic \ion{H}{i} in the field. For the high-resolution cross-correlation data, we have demonstrated that the feature is an artefact that appears in the data due to a significant increase in the receiver temperatures of the telescope array due to relatively bright \ion{H}{i} entering the main beam of the telescopes. 

The analysis of these EVN data has permitted us to develop and extend the analysis of VLBI data for the purposes of SETI and the detection of techno-signatures. VLBI, and interferometry more generally, offer advantages to SETI that single dishes and beam-formed arrays do not. We have also demonstrated that one must be careful in interpreting narrow band-signals even when they appear in VLBI cross-correlations. However, as the performance of software correlators continues to increase, and our capacity to analyse large data volumes also improves \citep{MA_AI}, VLBI can be an important technique for future SETI searches and in particular potential follow-up observations of candidate signals.

\section*{Acknowledgements}
Research reported in this publication was supported by a Newton Fund project, DARA (Development in Africa with Radio Astronomy), and awarded by the UK’s Science and Technology Facilities Council (STFC) - grant reference ST/R001103/1. This work made use of Astropy:\footnote{http://www.astropy.org} a community-developed core Python package and an ecosystem of tools and resources for astronomy \citep{astropy:2013, astropy:2018, astropy:2022}. This research has made use of the NASA Exoplanet Archive, which is operated by the California Institute of Technology, under contract with the National Aeronautics and Space Administration under the Exoplanet Exploration Program. This work has made use of data from the European Space Agency (ESA)
mission {\it Gaia} (\url{https://www.cosmos.esa.int/gaia}), processed by
the {\it Gaia} Data Processing and Analysis Consortium (DPAC,
\url{https://www.cosmos.esa.int/web/gaia/dpac/consortium}). Funding
for the DPAC has been provided by national institutions, in particular,
the institutions participating in the {\it Gaia} Multilateral Agreement.
$e$-MERLIN is a National Facility operated by the University of Manchester at Jodrell Bank Observatory on behalf of STFC. The European VLBI Network (www.evlbi.org) is a joint facility of independent European, African, Asian, and North American radio astronomy institutes. Scientific results from data presented in this publication are derived from the following EVN project code(s): RSG12. The research leading to these results has received funding from the European Commission Horizon 2020 Research and Innovation Programme under grant agreement No. 730562 (RadioNet).The research leading to these results has received funding from the European Commission Horizon 2020 Research and Innovation Programme under grant agreements No. 730562 (RadioNet) and No. 101004719 (OPTICON RadioNet Pilot).



\section*{Data Availability}
Data underlying this article are publicly available in the EVN Data Archive at JIVE at \url{www.jive.eu/select-experiment} and can be accessed with project codes RSG12 and RSW02. The $e$-MERLIN data and reduced EVN data will be shared on reasonable request to the corresponding author.
 



\bibliographystyle{mnras}
\bibliography{example} 








\bsp	
\label{lastpage}
\end{document}